\documentclass{workshop}
\usepackage{nohyperref}

\begin{document}

\title{
Early optical emission in support of synchrotron radiation in $\gamma$-ray bursts \\ 
}

\author{
Gor Oganesyan$^{1,}$$^{2,}$$^3$
 \\[12pt]  
$^1$ Gran Sasso Science Institute, Viale F. Crispi 7, I-67100, L’Aquila (AQ), Italy \\
$^2$ INFN - Laboratori Nazionali del Gran Sasso, I-67100, L’Aquila (AQ), Italy \\
$^3$ INAF - Osservatorio Astronomico d’Abruzzo, Via M. Maggini snc, I-64100 Teramo, Italy \\
\textit{E-mail: gor.oganesyan@gssi.it} 
}

\abst{
The origin of prompt emission in $\gamma$-ray bursts (GRBs) is highly debated topic. The observed spectra are supposed to play a crucial role in constraining the location of the emitting region, the strength of the magnetic field and the distribution of the accelerated particles. The apparent inconsistency of the prompt emission spectra with the synchrotron radiation scenario has resulted in considering more complex models. The inclusion of the soft X-ray data (down to 0.5 keV) in GRB spectra have led to the discovery of low-energy breaks in their spectra. More importantly, the distribution of spectral slopes has been shifted towards the prediction of the synchrotron radiation scenario if the break is associated with the synchrotron cooling frequency. We discuss the recent study that systematically extend the range of investigation down to the optical domain. It was shown that the optical-to-gamma-rays spectra are consistent with the synchrotron model. In addition, widely used empirical model made of thermal and non-thermal components has been tested. We conclude that most of the spectra are consistent with the synchrotron scenario while the two-component model faces difficulties to account for the optical radiation in presence/absence of the contaminating afterglow emission. We comment on the parameter space of GRB emitting region derived from the best fit parameters of the synchrotron model. In a basic one-shot particle acceleration model it corresponds to the quite contrived solutions for the magnetic field strength ($\sim$ 10 G) and for the radius of the emitting region ($R_\gamma \ge 10^{16}$\,cm). Possible modifications of the basic model would be necessary to have a fully consistent picture. 
}

\kword{workshop: proceedings --- gamma-ray bursts}

\maketitle
\thispagestyle{empty}

\section{Introduction}
The modern understanding of $\gamma$-ray bursts (GRBs) suggests that it originates from the internal dissipation of the kinetic and/or magnetic energy of a newly born and short-living relativistic jet. The observed variability, energetic and the spectral shapes of the prompt emission drive our theoretical understanding of the dissipation and radiation processes in the extreme jets of GRBs. The ultimate goal of any scrupulous analysis of GRB data is to make one more step towards our understanding of the jet formation mechanisms, its composition, energy transport, and efficiency and character of the particle acceleration. 

The most explored scenario for the GRBs is the hot fireball model \citep{1986ApJ...308L..43P,1986ApJ...308L..47G}. For the typical observed prompt emission luminosity $L \sim 10^{52}$~erg/s and the variability time-scale of $10^{-2}$~s, one gets an estimate the initial temperature of the ejecta as high as $T \sim  10^{10} $ (any units) guaranteeing photons, leptons and baryons being coupled. Jets with a small amount of baryons will undergo an adiabatic expansion, reaching bulk Lorentz factors of order of $\sim 100$ (see \citet{1990ApJ...365L..55S}). Due to the uncertainty on the composition and the energy transfer throughout the jet (the jets can be also dominated by the Poynting flux, see \citet{1992Natur.357..472U}), we do not understand the location of the emitting region (below or above the transparency radius) as well as the dominant mechanism for the jet's internal dissipation (shocks versus magnetic re-connection, for the wide range of references see \citet{2004RvMP...76.1143P,2015PhR...561....1K}). To re-construct back the physics of GRB jets, we can refer to the radiative processes that are shaping the observed spectra. The relative importance of the thermal and non-thermal components and their characteristics are then one of the main subjects of our interest. 

The observed spectra of GRBs in the $\sim10$~keV$-$10~MeV range are typically modelled as two power laws smoothly connected at a characteristic energy of hundreds of keV which corresponds to the peak energy in the $\nu F_{\nu}$ spectrum (e.g., \citet{1993ApJ...413..281B}). The presence of the well-established power-law tail above the peak energy and the overall inconsistency with the single black body spectrum indicates at first that the observed radiation is produced by a non-thermal population of charged particles rather then simply released from the photosphere of the pair-dominated fireball neither by the synchrotron radiation of the thermalized particles. The most straightforward model is then synchrotron radiation from the non-thermal population of electrons \citep{1994ApJ...430L..93R}. It was shown that the most efficient way to produce the prompt emission spectra in the keV-MeV range by the synchrotron radiation from the accelerated electrons requires the fast cooling regime, i.e., the electrons should cool at timescales much shorter than the dynamical time (e.g., see \citet{2000MNRAS.313L...1G}). The spectral shape of the fast cooling synchrotron emission below the peak energy has a value of -1.5 (photon index) independently from the injected electron's spectra. However, most of the measured spectral indices are larger than -1.5, with a typical value of -1 (e.g., see \citet{1998ApJ...506L..23P}). In other words, the GRB spectra are harder then expected in a synchrotron scenario. 

The fail of the simple synchrotron model to account for the GRB spectra has caused very intense theoretical and observational efforts in the literature to resolve the puzzle of the radiation process(es) responsible for GRBs. Some models have proposed modifications within the synchrotron model exploring the possibilities of including the marginally fast cooling regime, assymetry of the pitch angle distributions, cross-section dependent Inverse Compton effects, inhomogenous magnetic field. Other models have suggested photospheric radiation, sub-photospheric dissipation processes, Comptonisation, etc. (see \citet{2015PhR...561....1K} for a review). While different models are capable to explain the typical shapes or even the entire spectra, they correspond to quite different physical models for the GRB jets. Therefore, we are obliged to look for the non trivial ways of confrontation of the prompt emission models with the multi-wavelength and detailed spectral data. In the following, we briefly discuss the recent progresses in the spectral characterization of the GRB spectra by studying their low energy tails in the soft X-rays and optical bands. 

\section{Low-energy extension of the GRB spectra}
The inconsistency between the observed and the predicted spectral shapes in the standard fast cooling synchrotron radiation model lies on the low-energy part (below the peak energy) of the GRB spectra. Therefore, the characterization of the GRB spectra below the usual low-energy boundary of $\sim 10$~keV is a very promising tool for distinguishing the  proposed radiation models. 

\subsection{Prompt emission in soft X-rays}

The X-ray Telescope (XRT, 0.3-10 keV) on board of the Neil Gehrels Swift Observatory (hereafter, Swift; \citet{2004ApJ...611.1005G}) is a unique instrument allowing to partially cover the prompt emission of some long GRBs thanks to its rapid response and relatively fast slewing time to the GRB ($\sim 90$ s). The recent systematic study of the broad-band ($\sim$ 0.5 keV - 1 MeV) spectra of GRBs by an inclusion of XRT data has discovered a common feature of the low-energy spectral break at $\sim$ 2-20 keV \citep{2017ApJ...846..137O,2018A&A...616A.138O}. The spectra were shown to have a hardening below the break energy. Once the break is included in the analysis, the spectral indices below and above the break energy, on average, are consistent with the fast cooling synchrotron model if the break is associated with the synchrotron cooling frequency. These findings have motivated for the search of the similar spectral breaks at higher energies. \citet{2018A&A...613A..16R,2019A&A...625A..60R} have discovered the spectral breaks within the energy range of Fermi/GBM instrument (8 keV - 40 MeV) in the time-resolved spectra of the brightest GRBs. The spectral shapes below the peak energy, also in Fermi/GBM GRBs are found consistent with the synchrotron model in the marginally fast cooling regime. 

\subsection{Early optical emission}

The independent confirmation of the presence of the synchrotron-like breaks in the soft and hard X-ray range and by invoking different instrumentation has boosted our motivations to test the synchrotron model with early optical data. In the past, the synchrotron model alone has been rarely applied to the GRB spectra. \citet{1996ApJ...466..768T} has fitted the time-resolved spectra of few GRBs with the slow-cooling synchrotron model, while \citet{2000ApJ...543..722L} fitted two spectra of GRBs with the self-absorbed synchrotron model. Time-resolved spectra of GRB 130606B and several time-resolved spectra of GRB 160625B has been fitted satisfactorily by the single synchrotron model assuming the decay of the magnetic field within the emitting region \citep{2016ApJ...816...72Z,2018NatAs...2...69Z}. \citet{2019A&A...628A..59O} has performed the first systematic modelling of the GRB spectra by the single synchrotron radiation from a non-thermal population of electrons, taking into account their cooling\footnote{The first appearance of this result can be found in the PhD thesis, which can be downloaded from the SISSA website:  {http://hdl.handle.net/20.500.11767/84065}}. The results of this analysis has established that the synchrotron radiation alone is capable for accounting for the shapes of the time-resolved spectra of considered GRBs (21 long GRB, 52 spectra) when the cooling of electrons is taken into account. Moreover, the proper modelling of the GRB spectra by the physically derived synchrotron model has confirmed the marginally fast cooling regime of radiation indicated from the discovery of the low energy spectral breaks. Additionally, \citet{2019A&A...628A..59O} has proposed a novel approach to confront single synchrotron model to the widely applied two-component thermal plus non-thermal spectral model. They have used the early-time optical data as a tool to distinguish the two competing models. The basic idea lies on the extrapolation of the GRB spectral model at keV-MeV down to the optical bands. An acceptable model is required to predict the optical flux or at least to not overproduce it. They show that the synchrotron model successfully passes this trivial test, while the two-component model systematically overpredicts the optical flux contradicting the existence of the afterglow component. Moreover, the predicted flux from the synchrotron model is shown always to match the observed optical flux when the keV-MeV and the optical light curves are temporally correlated. And in other way around, when the optical light curves are in agreement with the radiation from an external shock, the synchrotron model savely underpredicts the level of the optical prompt emission. These series of tests give a strong preference to the single synchrotron radiation model for the production of the prompt emission. 

\section{Discussion}
The recent studies of the GRB spectra with the inclusion of the soft X-ray and optical data have found series of quite convincing arguments in support of the synchrotron radiation model to account for the production of the prompt emission. The marginally fast cooling regime suggested in these studies corresponds to a non trivial physical scenario. In a single-shot acceleration of electrons (a natural case in the internal shocks scenario), the marginally fast cooling regime returns large radii of the emitting regions ($R_\gamma \ge 10^{16}$\,cm), quite weak magnetic fields (order of unity in the comoving frame), large bulk Lorentz factors (few hundreds) and extreme energies of the accelerated electrons (Lorentz factors $\ge 10^{4}$) \citep{2008MNRAS.384...33K,2013ApJ...769...69B,2019A&A...628A..59O}. All of these constrains are in odds with our naive expectations from the GRB emission side: compact and highly magnetized regions located above the photosphere. Moreover, the requirement on large radii contradicts with the observed variability of the prompt emission within orders. Recently, these difficulties were discussed in details in \citet{2019arXiv191202185G}. They have shown that it is quite challenging to produce the observed marginally fast cooling regime by considering a non-thermal population of the electrons. They suggest a model with synchrotron radiation from protons as a possibility to overcome the basic difficulties that usual, electron-based models face to explain the observed prompt emission spectra. 

While we are still far from the complete understanding of the physics of the prompt emission, we can certainly conclude that optical and X-ray domains play critical role in discriminating and constraining the GRB models. The future wide field X-ray mission such as THESEUS \citep{2018AdSpR..62..191A} has a great potential to detect and characterise more in details the prompt emission in the soft X-rays \citep{2018MmSAI..89..245N}.

\label{last}

\end{document}